\documentclass{article}
\usepackage{amsmath,amssymb,amsthm,hyperref}
\usepackage[cm]{fullpage}

\newtheorem*{theorem}{The Nelson-Seiberg theorem revised}
\theoremstyle{remark} \newtheorem{remark}{Remark}

\begin{document}

\begin{figure}
\begin{flushright}
CAS-KITPC/ITP-339\\
CTP-SCU/2012001
\end{flushright}
\end{figure}

\title{The Nelson-Seiberg theorem revised}
\author{Zhaofeng Kang\textsuperscript{a, b, *},
        Tianjun Li\textsuperscript{a, c, d, \dag} and
        Zheng Sun\textsuperscript{a, e, \ddag}\\
        \textsuperscript{a}\normalsize\textit{State Key Laboratory of Theoretical Physics and Kavli Institute for Theoretical Physics China (KITPC),}\\
                           \normalsize\textit{Institute of Theoretical Physics, Chinese Academy of Sciences, Beijing 100190, P. R. China}\\
        \textsuperscript{b}\normalsize\textit{Center for High Energy Physics, Peking University, Beijing 100871, P. R. China}\\
        \textsuperscript{c}\normalsize\textit{George P. and Cynthia W. Mitchell Institute for Fundamental Physics and Astronomy,}\\
                           \normalsize\textit{Texas A \& M University, College Station, TX 77843, USA}\\
        \textsuperscript{d}\normalsize\textit{School of Physical Electronics, University of Electronic Science and Technology of China,}\\
                           \normalsize\textit{Chengdu 610054, P. R. China}\\
        \textsuperscript{e}\normalsize\textit{Center for Theoretical Physics, College of Physical Science and Technology,}\\
                           \normalsize\textit{Sichuan University, Chengdu 610064, P. R. China}\\
        \normalsize\textit{E-mail:}
        \textsuperscript{*}\texttt{zhfkang@itp.ac.cn,}
        \textsuperscript{\dag}\texttt{tli@itp.ac.cn,}
        \textsuperscript{\ddag}\texttt{sun\_ctp@scu.edu.cn}
       }
\date{}
\maketitle

\begin{abstract}

The well-accepted Nelson-Seiberg theorem relates R-symmetries to supersymmetry (SUSY) breaking vacua, and provides a guideline for SUSY model building which is the most promising physics beyond the Standard Model.  In the case of Wess-Zumino models with perturbative superpotentials, we revise the theorem to a combined necessary and sufficient condition for SUSY breaking which can be easily checked before solving the vacuum.  The revised theorem provides a powerful tool to construct either SUSY breaking or SUSY vacua, and offers many practicable applications in low energy SUSY model building and string phenomenology.

\end{abstract}

\section{Introduction}

The Standard Model (SM) is very successfully tested by current particle physics experiments.  It is completed after the recent discovery of a Higgs-like particle at the LHC.  But there still are a few clouds, for instance, the gauge hierarchy problem.  Supersymmetry (SUSY)~\cite{Martin:1997ns} provides a natural solution to these problems.  In the supersymmetric Standard Model (SSM) with an R-parity, gauge coupling unification can be realized, and the lightest supersymmetric particle (sparticle) becomes a dark matter candidate.  Since sparticles have not been discovered yet, SUSY must be broken to give them heavy masses escaping the current experimental limit.  To avoid light sparticles in the SSM, SUSY must be broken in a hidden sector, and then the SUSY breaking effects are transfered to the observable sector by a messenger sector, giving sparticle mass spectrum and coupling constants which may be examined at the LHC or other future experiments.

We consider the foundation of the above proposal, i.e., SUSY breaking in the hidden sector~\cite{Intriligator:2007cp}.  In model building, R-symmetries are often utilized because of their generic relation to F-term SUSY breaking vacua discovered by Nelson and Seiberg~\cite{Nelson:1993nf}.  The recent interest of metastable SUSY breaking~\cite{Intriligator:2006dd} also benefits from approximate R-symmetries~\cite{Intriligator:2007py, Abe:2007ax}.  Based on this, many dynamical SUSY breaking models with corresponding low energy effective description has been build and incorporated into phenomenology models.

Despite its great success, the original Nelson-Seiberg theorem has some defects which limit its scope of application.  First, it is stated as a necessary condition and a sufficient condition separately due to a singularity of the field redefinition used in the proof.  Second, the sufficient condition requires one to actually solve the vacuum, thus makes the theorem dispensable.  This work is to improve these issues and revise the theorem to a combined necessary and sufficient condition via a proof without field redefinitions.  The new proof is valid for any generic perturbative superpotentials.  As we are to see, the revised theorem can be easily checked before solving the vacuum, thus provides a powerful tool to construct either SUSY breaking or SUSY vacua, both of which are useful for model building in the hidden sector.

\section{The Nelson-Seiberg theorem revised}

Our setup is on a Wess-Zumino model which serves as a low energy effective description of many theories.  It contains a superpotential $W(\phi_i)$ which is a holomorphic function of chiral superfields $\phi_i, \ i = 1, \dotsc, d$ and a K\"ahler potential $K(\bar{\phi_i}, \phi_j)$ which is a real and positive-definite function.  Although a minimal K\"ahler potential $K(\bar{\phi_i}, \phi_j) = \sum_i \bar{\phi_i} \phi_i$ is often assumed in model building, most of our analysis (unless specified explicitly) is valid for generic K\"ahler potentials.  In the following notation, $W$ and $K$ are treated as functions of scalar components of $\phi_i$'s, namely $z_i$'s, because field values of $z_i$'s determine the vacuum structure of the model.  A vacuum corresponds to a minimum of the scalar potential
\begin{equation}
V = \sum_{i, j} K_{\bar{i} j} \partial_{\bar{i}} \bar{W} \partial_{j} W, \quad
K_{\bar{i} j} K^{\bar{i} j'} = \delta_j^{j'}, \quad
K^{\bar{i} j} = \partial_{\bar{i}} \partial_{j} K, \quad
\partial_{\bar{i}} = \partial_{\bar{z_i}}, \quad
\partial_i = \partial_{z_i}.
\end{equation}
Whether SUSY is broken or not can be checked by the F-term components $F_i = \partial_i W$ at the vacuum.  A solution to the equations $\partial_i W = 0$ corresponds to a SUSY vacuum which is also a global minimum of $V$.  As Nelson and Seiberg have pointed out~\cite{Nelson:1993nf}, SUSY solutions generically exist because there are equal numbers of equations and variables.  A non-R symmetry does not change the situation since it reduces both equations and variables by a same number.  The situation is different in an R-symmetric model where $W$ takes a special form because of the R-symmetry.  $W$ has R-charge $2$ in order to make the Lagrangian R-invariant.  One can select a field, $z_d$ supposedly, with a non-zero R-charge, and write $W$ with redefined fields as
\begin{equation} \label{eq:2-01}
W = x f(y_1, \dotsc, y_{d - 1}), \quad
x = z_d ^ {2 / r_d}, \quad
y_i = z_i / z_d ^ {r_i / r_d}, \quad
i = 1, \dotsc, d - 1,
\end{equation}
where $r_i$'s are R-charges of $z_i$'s.  For $x \ne 0$, the equations $\partial_i W = 0$ are equivalent to
\begin{equation}
f = 0, \quad
\partial_{y_i} f = 0, \quad
i = 1, \dotsc, d - 1.
\end{equation} 
There are $d$ equations and $d - 1$ variables.  So these equations can not be solved simultaneously for a generic function $f$.  A vacuum with $x \ne 0$, if existing, must be SUSY breaking in generic models.  The solution with $x = 0$ and $ f = 0$ could be a SUSY vacuum.  But the field redefinition in~\eqref{eq:2-01} is usually singular at $x = 0$, making the existence of such a vacuum unclear.  Notice that a non-zero expectation value of $x$ spontaneously breaks the R-symmetry, while at a vacuum with spontaneous R-symmetry breaking we can select a field $z_d \ne 0$ with $r_d \ne 0$ to define the needed $x$.  In summary, what we have presented is exactly the original Nelson-Seiberg theorem:  With generic parameters of $W$, an R-symmetry is a necessary condition, and a spontaneously broken R-symmetry is a sufficient condition for SUSY breaking at the global minimum of $V$.

The singular point of the field redefinition in~\eqref{eq:2-01} prevents us to discuss any solution at the origin.  This is the reason why the necessary and sufficient conditions are given separately in the theorem.  Furthermore, to utilize the sufficient condition, one has to solve a vacuum which avoids the singularity, thus the theorem becomes dispensable.  To overcome these defects, we present the R-symmetric $W$ without doing any field redefinition.  As shown in literature~\cite{Sun:2011fq}, fields can be categorized into three types according to their R-charges:
\begin{equation}
r(X_i) = 2, \quad
i = 1, \dotsc , N_X, \quad
r(Y_j) = 0, \quad
j = 1, \dotsc , N_Y, \quad
r(A_k) \ne 2 \text{ and } 0, \quad
k = 1, \dotsc , N_A.
\end{equation}
Considering a perturbative superpotential which can be expanded in a polynomial form, and keeping every term transforming correctly under the R-symmetry, we write down the generic form of $W$:
\begin{gather} \label{eq:2-02}
W = \sum_i X_i f_i(Y_j) + W_1,\\
\begin{split}
W_1 = &\sum_{\substack{i, j, k\\ r(A_k) = - 2}} \mu_{i j k} X_i X_j A_k
       + \sum_{\substack{i, j, k\\ r(A_j) + r(A_k) = 0}} \nu_{i j k} X_i A_j A_k
       + \sum_{\substack{i, j, k\\ r(A_j) + r(A_k) = 2}} \xi_{i j k} Y_i A_j A_k\\
      &+ \sum_{\substack{i, j\\ r(A_i) + r(A_j) = 2}} \kappa_{i j} A_i A_j
       + \sum_{\substack{i, j, k\\ r(A_i) + r(A_j) + r(A_k) = 2}} \lambda_{i j k} A_i A_j A_k
       + (\text{non-renormalizable terms}).
\end{split}
\end{gather} 
If $N_X \le N_Y$ is satisfied, all first derivatives of $W_1$ can be turned off by setting $X_i = A_i = 0$, and SUSY vacua can be found by solving $f_i(Y_j) = 0$.  Such a solution generically exists because the number of $f$'s is less than or equal to the number of $Y$'s.  One may try to rearrange R-charges to satisfy $N_X \le N_Y$ and get such SUSY vacua if there is arbitrariness in the R-charge assignment.  If $N_X > N_Y$ is always satisfied for any consistent R-charge assignment, generically there is no SUSY solution with $X_i = A_k = 0$ because there are always more $f$'s than $Y$'s.  In this case, if there is some $X_i \ne 0$ or $A_i \ne 0$, it breaks the R-symmetry spontaneously and ensures SUSY breaking via the original Nelson-Seiberg theorem.  Notice also that SUSY is generically unbroken without R-symmetries.  These exhaust all cases with and without R-symmetries.  Therefore we obtain the necessary and sufficient condition for SUSY breaking.

\begin{theorem}
In a Wess-Zumino model with a generic perturbative superpotential, SUSY is spontaneously broken at the global minimum if and only if the superpotential has an R-symmetry and there are more R-charge $2$ fields than R-charge $0$ fields for any consistent R-charge assignment.
\end{theorem}

Several remarks are to be addressed.

\begin{remark}
Renormalizability may cause non-genericness.
\end{remark}

Other than those non-generic exceptions where superpotential parameters take some special values as shown in literature~\cite{Nelson:1993nf}, renormalizability may cause another type of non-genericness because only up to cubic terms are allowed in a renormalizable $W$.  An example is the superpotential
\begin{equation} \label{eq:2-03}
W = f X + \lambda X A B, \quad
r_X = 2, \quad
r_A = 1 / 2, \quad
r_B = - 1 / 2.
\end{equation}
The R-charge assignment satisfies $N_X > N_Y$, but there is a SUSY vacuum at $X = 0$ and $A B = f / \lambda$, which also spontaneously breaks the R-symmetry for non-zero $f$ and $\lambda$.  Notice that with a minimal K\"ahler potential, there is a metastable SUSY breaking vacuum at $A = B = 0$ with $X$ being the pseudomodulus.  Adding to $W$ a non-renormalizable perturbation $\epsilon A^4$ (which respects the R-symmetry) destroys the SUSY vacuum and makes the model agreeing with both the original Nelson-Seiberg theorem and our revised one.

\begin{remark}
The arbitrariness in the R-charge assignment is equivalent to non-R $U(1)$ symmetries.
\end{remark}

As shown in literature~\cite{Komargodski:2009jf}, for a choice of R-charges $r_i$'s and non-R $U(1)$ charges $q_i$'s, R-charges can be reassigned as $r'_i = r_i + a q_i$ for $a \in \mathbb{R}$.  Non-R $U(1)$ charges can also be defined as $q_i = r'_i - r_i$ from two different choices of R-charges $r_i$'s and $r'_i$'s.  This arbitrariness is mentioned in the revised theorem and should be checked when utilizing the theorem.  An example is the previous superpotential~\eqref{eq:2-03} without non-renormalizable perturbations.  It has a non-R $U(1)$ symmetry with $q_X = 0$, $q_A = 1$ and $q_B = -1$, which can induce a new R-charge assignment $r_X = 2$ and $r_A = r_B = 0$.  Our revised theorem predicts the existence of the SUSY vacuum with such reassigned R-charges.  In addition, we point out that with a redefined field $Y = AB$, this model has effective $N_Y = 1$ and our theorem applies with such effective fields.  These argument of R-charge reassignment or effective fields may break down with non-renormalizable perturbations such as $\epsilon A^4$.

\begin{remark}
The existence of a global minimum should be taken as an assumption for the theorem.
\end{remark}

There are models with neither SUSY nor SUSY breaking vacuum, only maxima, saddle points and runaway directions.  An example is the superpotential
\begin{equation}
W = f X + \lambda X^2 A, \quad
r_X = 2, \quad
r_A = - 2,
\end{equation}
with a minimal K\"ahler potential.  There are a saddle point at $X = A = 0$ and a runaway direction along $A = f / (2 \lambda X)$ and $X \to 0$.  No vacuum exists in this model and even allowing non-renormalizable perturbations for $W$ does not help (but a non-minimal K\"ahler potential may help).  Such type of models are covered by neither the original Nelson-Seiberg theorem nor our revised one.  However runaway directions in such models may be lifted up, e.g., by non-vanishing D-terms~\cite{Azeyanagi:2012pc}, and provide useful vacua for model building.

\begin{remark}
The theorem may break down in the presence of non-perturbative effects.
\end{remark}

The generic form of the superpotential~\eqref{eq:2-02} is an essential step of our proof, which comes from the assumption of an R-symmetry and a perturbative superpotential.  A non-perturbative superpotential does not have a polynomial expansion like~\eqref{eq:2-02}, thus invalidates our proof.  The dynamical SUSY breaking $(3, 2)$ model~\cite{Affleck:1983mk} gives an example for this point.  One term in the superpotential proportional to $1 / Z$ is generated from an $SU(3)$ instanton, where $Z$ is a neutral composite field from product of several R-charged fields.  While there are no R-charge $2$ fields but several R-charge $0$ fields in the $(3, 2)$ model, a SUSY breaking vacuum is found at $Z \ne 0$ and the R-symmetry is also broken.  So this model provides a case within the scope of the original Nelson-Seiberg theorem but beyond our revised one.

\begin{remark}
With a minimal K\"ahler potential, the SUSY breaking pseudomodulus can not be a combination of only $Y$'s.
\end{remark}

The pseudomodulus, which makes a chiral multiplet with the goldstino and the F-term, exists on any metastable SUSY breaking vacuum in a Wess-Zumino model with a minimal K\"ahler potential~\cite{Ray:2006wk, Sun:2008nh}.  If $X$'s and $A$'s are not involved in the pseudomodulus, non-zero F-term components can only be some of $\partial_Y W$'s.  But from~\eqref{eq:2-02} we see varying $Y$'s along the pseudomodulus generates non-zero $\partial_X W$'s for generic $f$'s.  The contradiction proves our remark.  This remark means that a non-zero expectation value of the pseudomodulus always breaks the R-symmetry, no matter whether the SUSY breaking vacuum is a global minimum covered by our theorem or just a metastable local minimum.  It may hint that loop-level R-symmetry breaking~\cite{Shih:2007av, Curtin:2012yu} is more common than tree-level R-symmetry breaking~\cite{Komargodski:2009jf, Carpenter:2008wi, Sun:2008va}.

\section{Model building on SUSY breaking and SUSY vacua}

Our theorem provides a powerful tool to construct either SUSY breaking or SUSY vacua in generic models.  One can easily arrange R-charges of fields satisfying either $N_X > N_Y$ or $N_X \le N_Y$ to get the needed vacua.  Since SUSY breaking vacua are rare compared to SUSY vacua in models without R-symmetries even considering metastable vacua~\cite{Sun:2011aq}, the help of R-symmetries through the theorem is welcome for SUSY breaking model building.  One can start from an R-symmetric model with $N_X > N_Y$.  The global minimum, if existing, is a SUSY breaking vacuum.  Adding R-symmetry breaking terms to $W$ can restore SUSY vacua.  But one may expect that SUSY breaking vacua retain their local metastability and are long-lived against the Coleman-de~Luccia decay~\cite{Coleman:1977py, Coleman:1980aw} if the R-symmetry breaking is small enough.  This is the general procedure of many constructions of metastable SUSY breaking models with approximate R-symmetries~\cite{Intriligator:2007py, Abe:2007ax}.  One may then seek for dynamic SUSY breaking models, such as the ones from Seiberg duality~\cite{Seiberg:1994pq}, which have corresponding effective Wess-Zumino descriptions at low energy.  As a bonus, the generic form of the superpotential~\eqref{eq:2-02} allows arbitrary number of $A$'s with R-charges other than $2$ and $0$, which do not alter the statement of our theorem.  These $A$'s are essentially required for spontaneous R-symmetry breaking~\cite{Shih:2007av, Curtin:2012yu, Sun:2008va} which generates the masses of one type of sparticles, i.e., gauginos.

As for model building on SUSY vacua, the generic form of the superpotential~\eqref{eq:2-02} shows that satisfying $N_X \le N_Y$ provides the bonus $W = 0$, which leads to SUSY vacua with zero cosmological constants in the supergravity (SUGRA) extension of the Wess-Zumino model.  This part of the result is also true for non-$\mathbb{Z}_2$ discrete R-symmetries~\cite{Sun:2011fq}.  These properties grant such vacua an important role in string phenomenology.  One may identify a discrete R-symmetry to one of the geometrical symmetries of the compact manifold when compactifying string theory from $10$ to $4$ dimensions.  In the framework of flux compactifications, $N_X \le N_Y$ can be easily arranged in many models with a low energy effective SUGRA description, producing a huge landscape of SUSY vacua with zero cosmological constants at tree level~\cite{Dine:2005gz}.  SUSY is then broken dynamically on these vacua through non-perturbative corrections to generate an exponentially small scale.  By this means the hierarchy from the Planck scale to the scale of low energy SUSY or the cosmological constant may be understood naturally, and the real-world physics such as the SM and SSM may hopefully be built on these vacua from string theory.

A key improvement of our revised theorem compared to the original one is the convenience of the criterion $N_X > N_Y$ or $N_X \le N_Y$.  Assuming genericness of parameters, utilizing our theorem does not require the explicit form of $W$.  This allows one to efficiently survey a vast set of different models and select the desired one to continue the explicit model building.  The question of SUSY breaking versus SUSY or the prediction of the SUSY breaking scale may also be answered by employing this procedure in the landscape of string theory or other fundamental theories.  In summary, our revised theorem offers many practicable applications in low energy SUSY model building as well as string phenomenology.

\section*{Acknowledgement}

We thank Feihu Liu for helpful discussions.  This work is supported by the National Natural Science Foundation of China under grant 10821504, 11075194, 11135003 and 11305110, by the Project of Knowledge Innovation Program (PKIP) of Chinese Academy of Sciences under grant KJCX2.YW.W10, and by the DOE grant DE-FG03-95-Er-40917.

\end{document}